\documentclass[
 reprint,
superscriptaddress,
showkeys,
 amsmath,amssymb,
 aps,
]{revtex4-2}

\usepackage{graphicx}% Include figure files
\usepackage{dcolumn}% Align table columns on decimal point
\usepackage{bm}% bold math
\usepackage{hyperref}% add hypertext capabilities
\usepackage{xcolor}
\usepackage{titlecaps}

\begin{document}

\preprint{APS/123-QED}

\title{A Polymer Chain with Dipolar Active Forces in Connection to Spatial Organization of Chromatin}

\author{Subhasish Chaki}
\email{subhasischaki@gmail.com}
\affiliation{Department of Materials Science and Engineering, University of Illinois, Urbana, Illinois - 61801, USA}
\affiliation{Department of Chemistry, Indian Institute of Technology Bombay, Mumbai, Maharashtra -  400076, India}
\author{Ligesh Theeyancheri}
\email{ligeshbhaskar@gmail.com}
\affiliation{Department of Chemistry, Indian Institute of Technology Bombay, Mumbai, Maharashtra -  400076, India}
\author{Rajarshi Chakrabarti}
\email{rajarshi@chem.iitb.ac.in}
\affiliation{Department of Chemistry, Indian Institute of Technology Bombay, Mumbai, Maharashtra -  400076, India}

\begin{abstract}
\noindent A living cell is an active environment where the organization and dynamics of chromatin are affected by different forms of activity. Optical experiments report that loci show subdiffusive dynamics and the chromatin fiber is seen to be coherent over micrometer-scale regions. Using a bead-spring polymer chain with  dipolar active forces, we study how the subdiffusive motion of the loci generate large-scale coherent motion of the chromatin. We show that in the presence of extensile (contractile) activity, the dynamics of loci grows faster (slower) and the spatial correlation length increases (decreases) compared to the case with no dipolar forces. Hence, both the dipolar active forces modify the elasticity of the chain. Interestingly in our model, the dynamics and organization of such dipolar active chains largely differ from the passive chain with renormalized elasticity.
\end{abstract}

%\keywords{} 

\maketitle
%%%%%
\section{Introduction}

\noindent Active systems are inherently out of equilibrium \cite{gnesotto2018broken}. It consumes energy perpetually from the environment which prevents relaxation into a thermal equilibrium state and exhibits a wide spectrum of novel steady-state behaviours such as the activity-driven phase separation \cite{smrek2017small} or large-scale collective motion \cite{schaller2010polar}. Even inside the living cells, active systems have constituents that adapt in noisy environments and generate forces for uni-directional transport and spatial organization. Biopolymers are an integral part of living cells, and their conformational and dynamical properties are significantly affected by the active processes \cite{winkler2020physics,osmanovic2017dynamics,samanta2016chain,chaki2019enhanced,joshi2019interplay,chelakkot2014flagellar,shin2015facilitation,goswami2022reconfiguration}. For example, the microtubule and F-actin filaments of the cytoskeleton are driven forward by the myosin and kinesin families of motor proteins \cite{le2001motor,brangwynne2008nonequilibrium}. Another biopolymer inside the nucleus is chromosomal DNA \cite{maji2020lamin,bajpai2020irregular,salari2022spatial} and it serves as a platform for number of biological processes such as Chromatin remodelling \cite{natesan2021active}, transcription \cite{andersson2015unified} and DNA repair \cite{eaton2020structural}. Their spatial organization dramatically varies along with the different phases of the cell cycle, which seem to be driven by processes that consume ATP \cite{agrawal2017chromatin,maharana2012dynamic}. However, the exact procedure of energy consumption on the chromatin remains unclear and this is the primary interest of the present study. \\

\noindent Interphase chromatin are organized into transcriptionally active and inactive regions. Transcriptionally active regions, referred to as euchromatin, are loosely packed, whereas transcriptionally inactive regions, called heterochromatin, are tightly packed \cite{papantonis2013transcription}, \cite{shi2018interphase}. In a recent study, transcriptionally active regions are characterized by an effective temperature in which a Brownian-like random force is acting on chromosomes in addition to thermal noise \cite{agrawal2020nonequilibrium,chubak2022active}.  Weber $et. al.$  argued that random motions of chromatin loci in $E.$ $Coli$ are driven by the ATP-dependent fluctuations which behave like thermal fluctuations but with a greater magnitude \cite{weber2012nonthermal}. However, Liu $et.\,\, al.$ showed that the enhanced thermal like fluctuations due to activity randomize the global structure of chromatin chain more efficiently as compared to the thermal noise and shorten the spatial correlation at a sufficiently large time \cite{liu2018chain}. On contrary, Zidovska $et.\,\,al.$ demonstrated two types of chromatin motion experimentally: a fast local motion previously observed by tracking the motion of chromosomal loci and a slower large-scale motion in which collective motions of chromatin are seen to be coherent beyond the boundaries of chromosome territories, over micrometer-scale regions and seconds \cite{zidovska2013micron}. To explain the correlated motions of chromatin, Saintillan $et.\,\,al.$  constructed a flexible polymer model with long-range hydrodynamic interactions that is driven by active force dipoles \cite{saintillan2018extensile}. In their model, they obtained strong correlated motion for extensile activity, but not in the cases of passive and contractile activity \cite{mahajan2022self}. In another study, Liu $et.\,\,al.$  modelled the chromatin as active flexible polymer subjected to extensile and contractile monopoles \cite{liu2021dynamic}. They  showed  that both the extensile and contractile monopoles generate correlated motion without the hydrodynamic interactions and the contractile system exhibits  larger displacements than the extensile one. Other theoretical models of active polymers have also been proposed in connection to chromosomal dynamics. Some of these models have focused on semiflexible \cite{ghosh2014dynamics} or flexible polymer \cite{put2019active} chains subject to correlated noise  or an active force parallel to the backbone of the chain \cite{bianco2018globule}. Hence, the dynamics is superdiffusive in such models. A recent study in mouse embryonic stem cells showed that  differentiated chromatin moves by an apparent free diffusion, while the undifferentiated one behaves superdiffusively \cite{eshghi2021interphase}. \\

\noindent To understand how the large-scale correlated chromatin dynamics emerges from the local subdiffusive motion, we construct a model for chromatin as a long semiflexible polymer with excluded volume interactions and dipolar extensile and contractile activities. We do not include hydrodynamic interactions in this study. We show that both the extensile and contractile active forces modify the elasticity of the chain, but leave the subdiffusive nature of the loci dynamics unchanged. We compare our results for dipolar activity to thermal system with renormalized elasticity and find that dipolar active forces create larger expansion or compaction of the chain as compared to thermal system with renormalized elasticity. Hence,  dipolar active forces can not be thought of as  a thermal system with renormalized elasticity.  \\

\section{Model}

\begin{figure*}
\centering
	\includegraphics[width=0.9\linewidth]{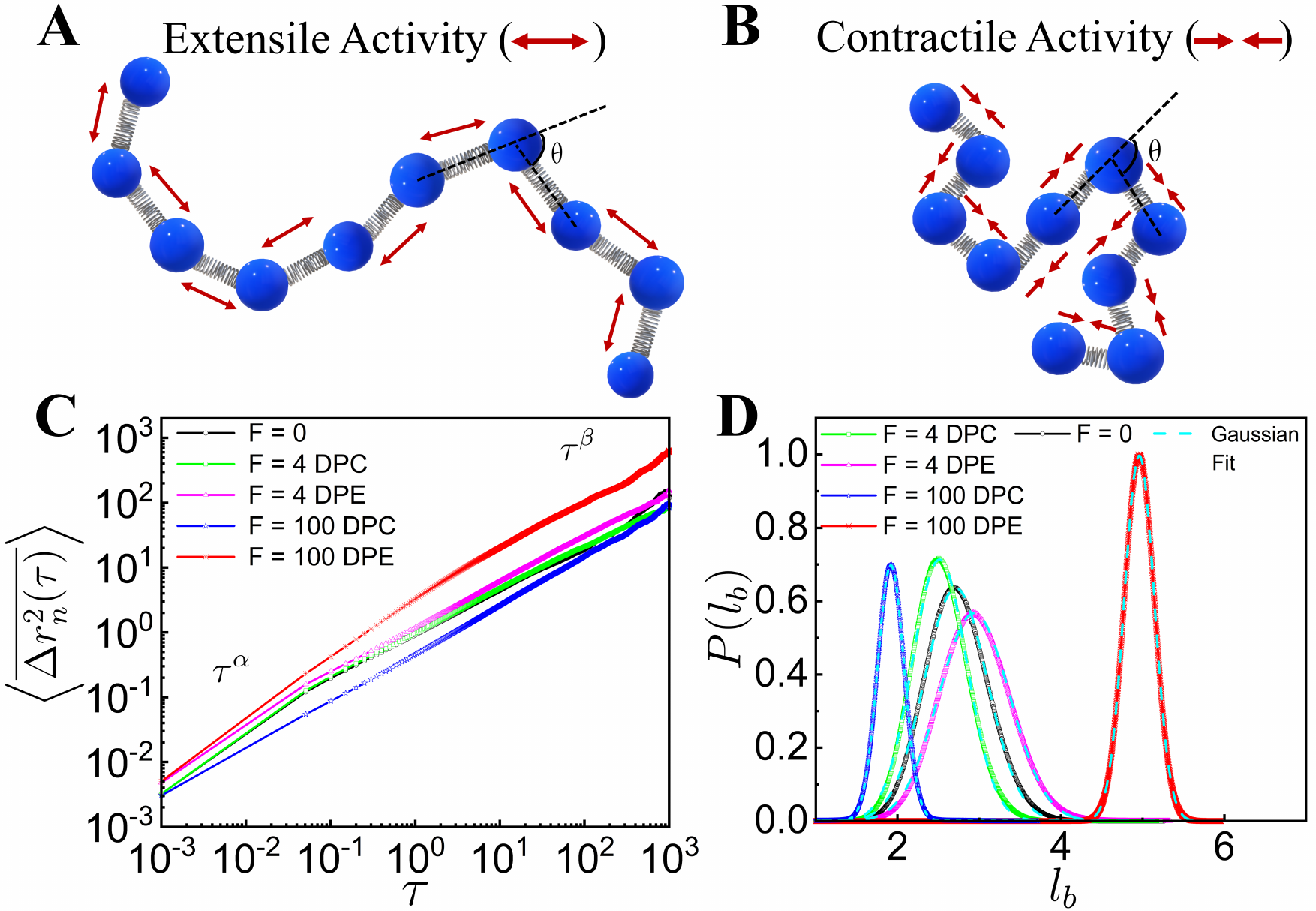} 
\caption{A schematic representation of the semiflexible polymer chain subjected to the extensile (A) and contractile (B) motor activities. $\theta$ is the bond angle in both the figures. (C) Log-Log plot of tagged monomer MSDs $\left(\left<\overline{\Delta r_n^2(\tau)}\right>\right)$ as a function of lag time $\tau$ and  (D) the probability distribution functions, $P(l_b)$ of the bond length, $l_b$  at different DPE and DPC active forces.}\label{fig:MSD_bond_length_schematic}
\end{figure*}

\noindent We consider a bead-spring  polymer chain composed of $1000$ monomers of radius $\sigma$ in a three-dimensional space. We implement the following overdamped Langevin equation to simulate the dynamics of all the monomers of our system with the position $r_{n}$(t) at time $t$:
\begin{equation}
	\xi \frac{d \vec{r}_{n}}{dt}=- \nabla_n V(\vec{r}_{1},\vec{r}_{2}........)+ {\vec f}_{n}(t)+{\vec F}_{a,n},
	\label{eq:langevineq}
\end{equation} 
where  $\xi $ is the friction coefficient which is related to the thermal Gaussian noise through
\begin{equation}
	\left<{\vec f}_{n}(t)\right>=0, \hspace{5mm}
	\left<{\vec f}_{n}(t^{\prime}).{\vec f}_{m}(t^{\prime\prime})\right> =6 \xi k_B T \delta_{nm}\delta(t^{\prime}-t^{\prime\prime}).
	\label{eq:random-forcerouse}
\end{equation}
The total potential energy, $V(\vec{r}_{1},\vec{r}_{2}........) = V_{\text{angle}}+V_{\text{WCA}}+V_{\text{bond}}$ is composed of a bond contribution between neighbouring beads
\begin{equation}
V_{\text{bond}}= \frac{k}{2} \left(|\vec{r}_{n,n+1}|-r_0\right)^2, 
\label{eq:harmonic}
\end{equation}
a bending energy
 \begin{equation}
V_{\text{angle}}\left(\theta_i \right)=k_a \left(1-\cos\theta_i\right),
\label{eq:angle}
\end{equation}
and an excluded volume interaction modelled with Weeks-Chandler-Andersen (WCA) potential \cite{weeks1971role}:
\begin{equation}
V_{\textrm{WCA}}(r)=\begin{cases}4\epsilon\left[\left(\frac{\sigma}{r}\right)^{12}-\left(\frac{\sigma}{r}\right)^{6}\right]+\epsilon, \mbox{if }r<2^{1/6}\sigma \\
0, \hspace{35mm} \mbox{otherwise},
\end{cases}
\label{eq:WCA}
\end{equation}
where $\vec{r}_{n,n+1}=\vec{r}_{n}-\vec{r}_{n+1}$ is the vector between the position of the beads $n$ and $n+1$, $k$ is the force constant, $r_0$ is the equilibrium bond distance, $\cos\theta_n=\hat{r}_{n,n+1}(t).\hat{r}_{n+1,n+2}(t)$, $k_a$ is the  bending modulus, $r$ is the separation between a pair of monomers in the medium, $\epsilon$ is the strength of the interaction, and $\sigma$ determines the effective interaction diameter. Without the force, ${\vec{F}}_{a,n}=0$, the model matches with the well-known worm-like chain model for semi-flexible polymers. Here ${\vec{F}}_{a,n}$ denotes the dipolar active forces on each bead that pushes simultaneously two neighbouring beads towards (contractile) or away (extensile) from each other along the unit bond vector for all the time with constant magnitude $F$ (Fig. \ref{fig:MSD_bond_length_schematic} (A, B)). For the extensile case, the force $F \frac{\vec{r}_{n+1}-\vec{r}_{n}}{|\vec{r}_{n+1}-\vec{r}_{n}|}$ and $F \frac{\vec{r}_{n-1}-\vec{r}_{n}}{|\vec{r}_{n-1}-\vec{r}_{n}|}$ are exerted on the $(n+1)^{\textrm{th}}$ and $(n-1)^{\textrm{th}}$ monomers respectively. Thus, the extensile force on the $n^{\textrm{th}}$ monomer, ${\vec F}_{a,n} =F\left[ \frac{\vec{r}_{n}-\vec{r}_{n+1}}{|\vec{r}_{n}-\vec{r}_{n+1}|}+ \frac{\vec{r}_{n}-\vec{r}_{n-1}}{|\vec{r}_{n}-\vec{r}_{n-1}|}\right]$. Similarly, for the contractile force, ${\vec F}_{a,n} =F\left[ \frac{\vec{r}_{n+1}-\vec{r}_{n}}{|\vec{r}_{n+1}-\vec{r}_{n}|}+ \frac{\vec{r}_{n-1}-\vec{r}_{n}}{|\vec{r}_{n-1}-\vec{r}_{n}|}\right]$. This is where our model differs from the existing models where activity is introduced by putting the beads at a higher temperature \cite{agrawal2020nonequilibrium,chubak2022active}. The dipolar extensile or contractile active forces in our model can be derived from a potential $V_{a,n}=-F \left(|\vec{r}_{n+1}-\vec{r}_{n}|+ |\vec{r}_{n}-\vec{r}_{n-1}|\right)$ or $V_{a,n}=F \left(|\vec{r}_{n+1}-\vec{r}_{n}|+ |\vec{r}_{n}-\vec{r}_{n-1}|\right)$. In the works by Prathyusha $et.$ $al.$ \cite{prathyusha2018dynamically} or Isele-Holder $et.$ $al.$ \cite{isele2015self},  the magnitude of the active force is constant and the direction is acting along the unit bond vector of the polymer. The active forces modeled in these works can also be generated from a potential, but not dipolar in nature. Therefore, the composite system made of polymer and the activity behave like a thermal system and the total force on the composite system will be conservative. We do not incorporate any binding rates of the active force  \cite{ghosh2014dynamics, put2019active} in the model which might modify the magnitude of the active force by a factor. In our simulation, we consider $\sigma$, $k_B T$ and $\tau_0=\frac{\sigma^2 \xi}{k_B T}$ as the unit of length, energy and time scales. The parameters used for the simulation are $k=30 k_B T/\sigma^2$, $r_0=2 \sigma$, $k_a=2 k_B T$, $\epsilon=k_B T$. All the production simulations are carried out for $10^7$ steps where the integration time step is considered to be  $\delta t=10^{-5}\tau_0$. The simulations are carried out using LAMMPS \cite{plimpton1995fast}, a freely available open-source molecular dynamics package. For a given set of  parameters, we generate $10$ independent trajectories of the polymer.\\

\section{Results}

\subsection{Mean Square Displacement}

\noindent To quantify the effect of dipolar extensile (DPE) and dipolar contractile (DPC) activities on the dynamics of the individual monomers, we first focus on the time-and-ensemble averaged mean squared displacement (MSD) of the tagged monomer in the reference frame of the polymer's center of mass
(COM), $\left\langle{\overline{\Delta r_{n}^{2}(\tau)}}\right\rangle$  as a function of lag time $\tau$. The time-averaged MSD of $n$-th monomer is defined as $\overline{\Delta r_n^{2}(\tau)} = \frac{1}{T-\tau} \int_{0}^{T-\tau} {\left[ \vec{r}_n(t_0+\tau) - \vec{r}_n(t_0)\right]}^2  dt_0$ from the time series $\vec{r}_n(t)$ where $T$ is the total run time. The time-and-ensemble-averaged MSD is obtained by summing over all the monomers as: $\left\langle{\overline{\Delta r_n^{2}(\tau)}}\right\rangle  =  \frac{1}{N} \sum_{n=1}^{N}{\overline{\Delta r_n^{2}(\tau)}}$  where $N$ is the number of monomers. For small DPE and DPC activities $(F=4)$, the dynamics of the tagged monomer is subdiffusive (see supplementary Table S1) and we do not see any significant enhancement in $\left<\overline{\Delta r_n^2(\tau)}\right>$ as compared to the passive case $(F=0)$ as evident in Fig. \ref{fig:MSD_bond_length_schematic}C. But, for very high DPE and DPC activities $(F=100)$, we observe a clear difference in $\left<\overline{\Delta r_n^2(\tau)}\right>$, while the subdiffusive dynamics of the monomer remains unaltered. In comparison to the passive case, the dynamics of the individual monomer is enhanced for high DPE force and suppressed for high DPC force respectively (Fig. \ref{fig:MSD_bond_length_schematic}C).  However, the dipolar activity leaves the dynamical behavior of the COM unchanged (see supplementary Fig. S1), suggesting that the effective temperature as introduced in earlier studies of chromatin  \cite{agrawal2020nonequilibrium,ghosh2014dynamics}, is not a good descriptor for dipolar activity. Hence, the responses of the COM and individual monomer in the presence of dipolar activity suggest that the dipolar forces  modify the stretching of the chain which essentially enhances or reduces the local space explored by the tagged monomer for the DPE or DPC forces respectively, but not  the diffusive or subdiffusive nature of the displacements. \\

\subsection{Distribution of Bond Lengths}

\noindent To further quantify the effects of the DPE and DPC active forces on the loci mobilities, we investigate the probability distribution, $P(l_b)$ of the bond length, $l_b (=|\vec{r}_n-\vec{r}_{n-1}|)$ of the chain. We calculate $l_b$ for $N-1$ different bonds at each time-step of every simulation after the system reaches the steady-state where $N$ is the number of monomers. Then a single trajectory is created by joining different individual trajectories.  This single trajectory is binned to construct a histogram from which the ensemble averaged probability distribution $P(l_b)$ is computed. Here  $P(l_b)$ values follow Gaussian distributions (Fig.  \ref{fig:MSD_bond_length_schematic}D). For smaller DPC and DPE activities, $P(l_b)$ values appear to be qualitatively similar to $P(l_b)$ of the passive system. But, $P(l_b)$ values for high DPE and DPC forces show well-separated peaks (Fig.  \ref{fig:MSD_bond_length_schematic}D). For large DPE force, the peak of $P(l_b)$ shifts towards a higher value of $l_b$ which accounts for the enhancement of stretching of the polymer chain. For large contractile force, $P(l_b)$ is sharply peaked at lower value of $l_b$ and the deviation of the peak value of $P(l_b)$ from the passive case is small as compared to the deviation for the DPE force. This small deviation for the contractile force is presumably due to the self-avoidance of the beads. The potential energy (PE) per particle increases more for the DPC forces than for the DPE forces (see supplementary Fig. S2A) as the effect of self-avoidance translates to an increasing potential energy barrier for the contractile forces. On the other hand, the mean square fluctuation of the velocity $(v)$, $\left<v^2\right>$ per particle increases more for the DPE forces than for the DPC forces (see supplementary Fig. S2B) as the DPE forces show larger displacements than the DPC forces. Additionally, the probability distribution, $P(R_g)$ of the radius of gyration $R_g$ of the chain and the distribution, $P(N)$ of number of monomers, $N$ in a spherical probe volume indicate that DPE (or DPC) forces promote larger (or smaller) displacements of the system than the passive one (see supplementary Sec. III \& IV and Fig. S3 \& S4). \\ 

\subsection{Spatio-temporal Correlation}

\begin{figure*}
\centering
	\includegraphics[width=0.9\linewidth]{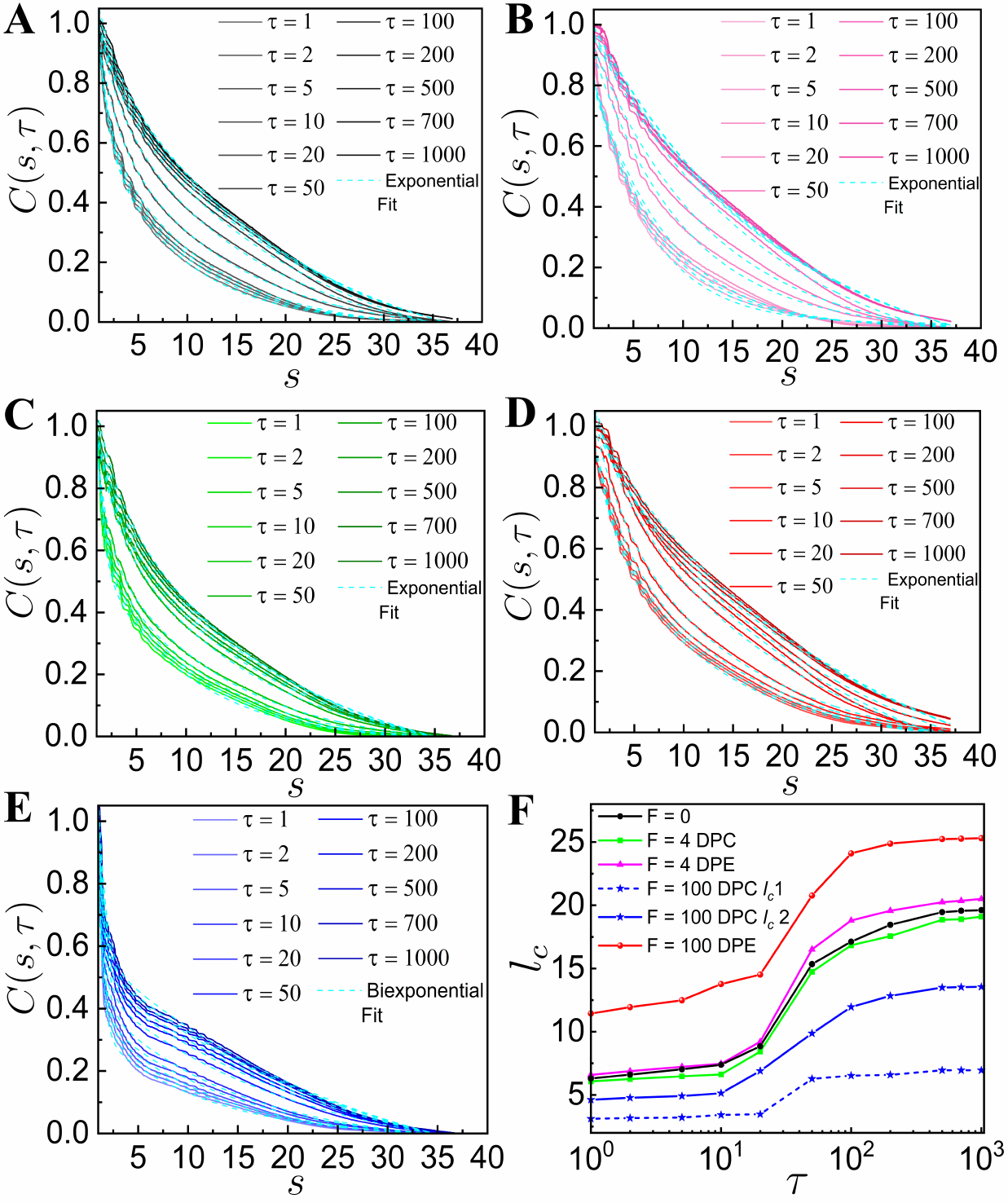} 
\caption{Spatial correlation functions $C(s,\tau)$ for the passive (A), DPE and DPC cases for F = 4 (B and C), F = 100 (D and E) respectively  at different time lags, $\tau$. (F) Correlation lengths ($l_c$) as a function of $\tau$ for the passive, DPE and DPC activities. }\label{fig:spatial}
\end{figure*}

\noindent To study the spatio-temporal dynamics of our chromatin model, we calculate the spatial correlation function, $C(s,\tau)$ as a function of spatial distance between two loci for different values of the lag time $\tau$ \cite{zidovska2013micron}. The normalized spatial correlation between $i$-th and $j$-th loci over the time interval $\tau$ is evaluated as $C(s,\tau)=\left<\left<\frac{\sum_{i>j}^N \left[ \vec{r}_i(t+\tau) - \vec{r}_i(t)\right] . \left[ \vec{r}_j(t+\tau) - \vec{r}_j(t)\right] \delta(s-|\vec{r}_i(t)-\vec{r}_j(t)|)}{\sum_{i>j}^N  \delta(s-|\vec{r}_i(t)-\vec{r}_j(t)|)}\right>_{t}\right>,$ where $\left<\left<....\right>_{t}\right>$ is an average over $t$ and over different ensembles. As $\tau$ increases, $C(s,\tau)$ values show slower spatial decay both in the case of with and without the active dipolar forces (Fig.  \ref{fig:spatial}). For passive and smaller DPC and DPE activities, $C(s,\tau)$ values decay exponentially (Fig.  \ref{fig:spatial} (A-C)). The corresponding correlation lengths, $l_c$  increase monotonically with $\tau$ and eventually saturated at large $\tau$ (Fig.  \ref{fig:spatial}F), which is consistent with the experimental observation \cite{shaban2018formation}. $C(s,\tau)$ shows similar exponentially decaying correlation with $\tau$ for higher DPE  force (Fig.  \ref{fig:spatial}D). For large DPC force,  we fit $C(s,\tau)$ with an bi-exponentially decaying function where $l_{c_1}(\tau)$ and $l_{c_2}(\tau)$ represent the correlation lengths for the initial faster and subsequent slower decay respectively (Fig.  \ref{fig:spatial} (E, F)). The correlation length $l_c$ of  extensile  active systems is greater than the same for the contractile active systems. This suggests that the motion of  extensile active chain has long-ranged spatial coherence than the contractile active chain. \\

\begin{figure*}
        \centering
        \includegraphics[width=0.9\linewidth]{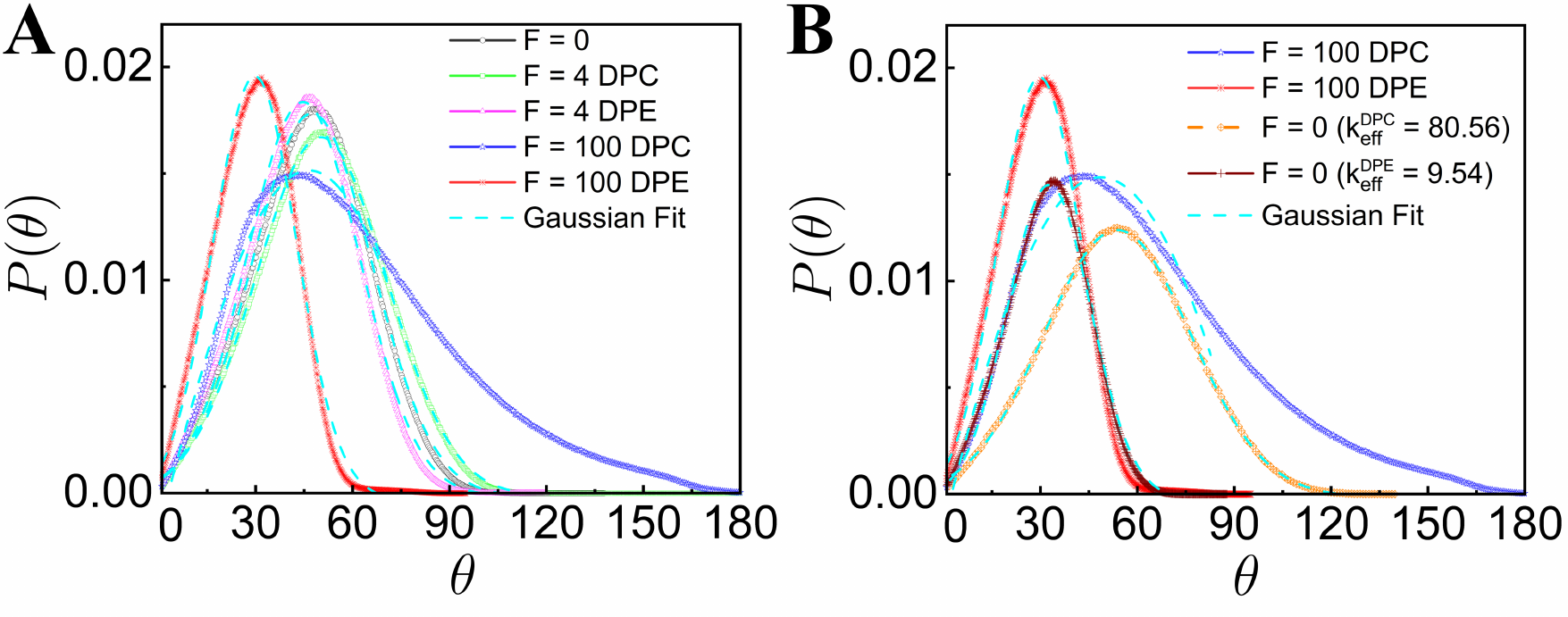}
        \caption{Probability density functions of the local curvature,  $P(\theta)$ for (A) passive, DPE and DPC activities and (B) for the passive chain with modified spring constants, $k^{\textrm{DPE}}_{\textrm{eff}}$ and $k^{\textrm{DPC}}_{\textrm{eff}}$ and large DPE and DPC dipolar active forces.}\label{fig:bending_keff}
\end{figure*}

\subsection{Distribution of Bending Fluctuations}

\noindent Other than $C(s,\tau)$, we look at the normalized distribution of local curvatures $P(\theta)$ to characterize the internal structure more quantitatively, where $\theta$ is the bond angle between the bond vectors.  Fig. \ref{fig:bending_keff}A shows that for small DPC and DPE activities, $P(\theta)$ values are Gaussian and nicely collapse on $P(\theta)$ of the passive system. However, for large DPC and DPE systems, $P(\theta)$ values display a clear difference from passive system.  $P(\theta)$ for DPE chain with higher activity exhibits a peak at smaller bond angle and the width of $P(\theta)$ becomes narrower as compared to that of the  DPC chain with higher activity (Fig. \ref{fig:bending_keff}A). Fig. \ref{fig:bending_keff}B  is discussed in the latter part of the manuscript. Narrowed distribution of $P(\theta)$ with peak value at small bond angle for high DPE force implies effectively higher bending rigidity as compared to that of passive chain. This again confirms that extensile activity induces long-ranged spatial correlations compared to contractile activity. \\

\begin{figure*}
\centering
	\includegraphics[width=0.9\linewidth]{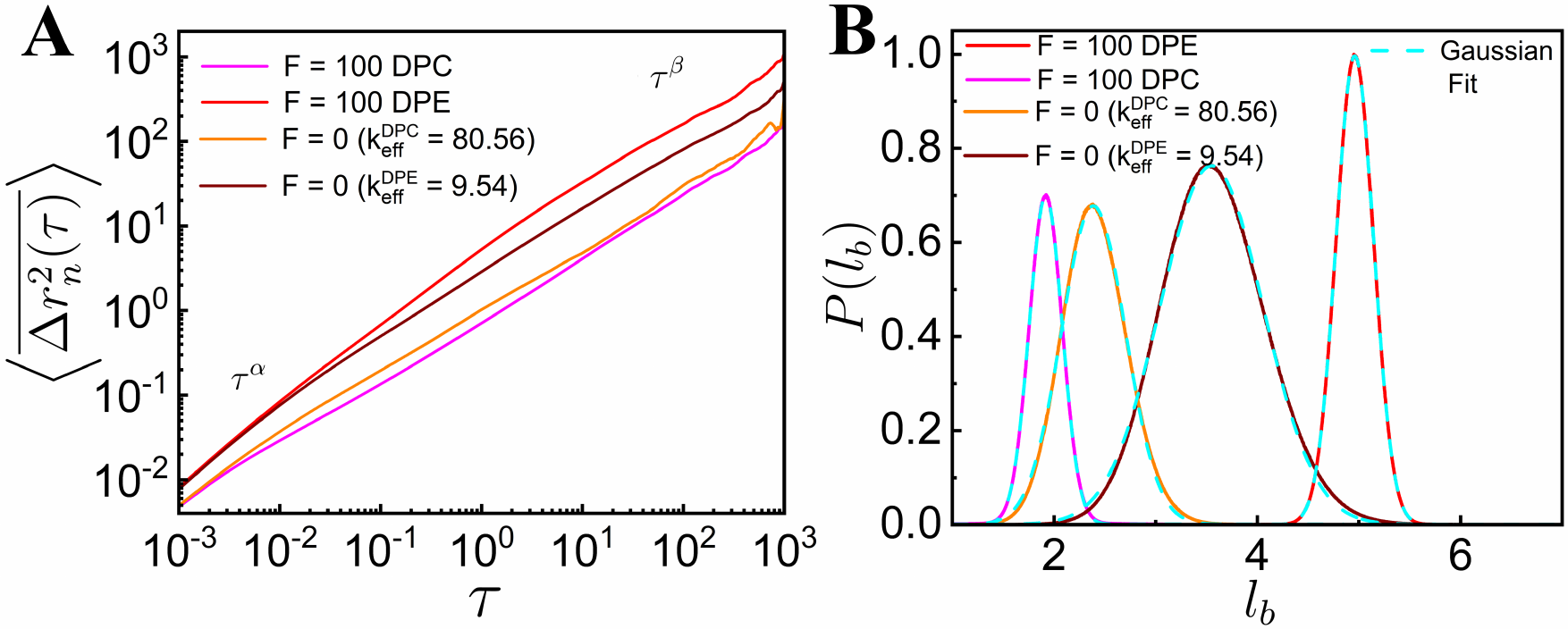} 
\caption{(A) log-log plot of $\left<\overline{\Delta r_n^2(\tau)}\right>$ as a function of lag time $\tau$  and (B) Probability density functions of bond lengths ($P(l_b)$)  for the thermal chain with renormalized spring constants, $k^{\textrm{DPE}}_{\textrm{eff}}$ and $k^{\textrm{DPC}}_{\textrm{eff}}$ and large DPE and DPC dipolar active forces.}\label{fig:msd_bond_modk}
\end{figure*}

\begin{figure*}
\centering
	\includegraphics[width=0.9\linewidth]{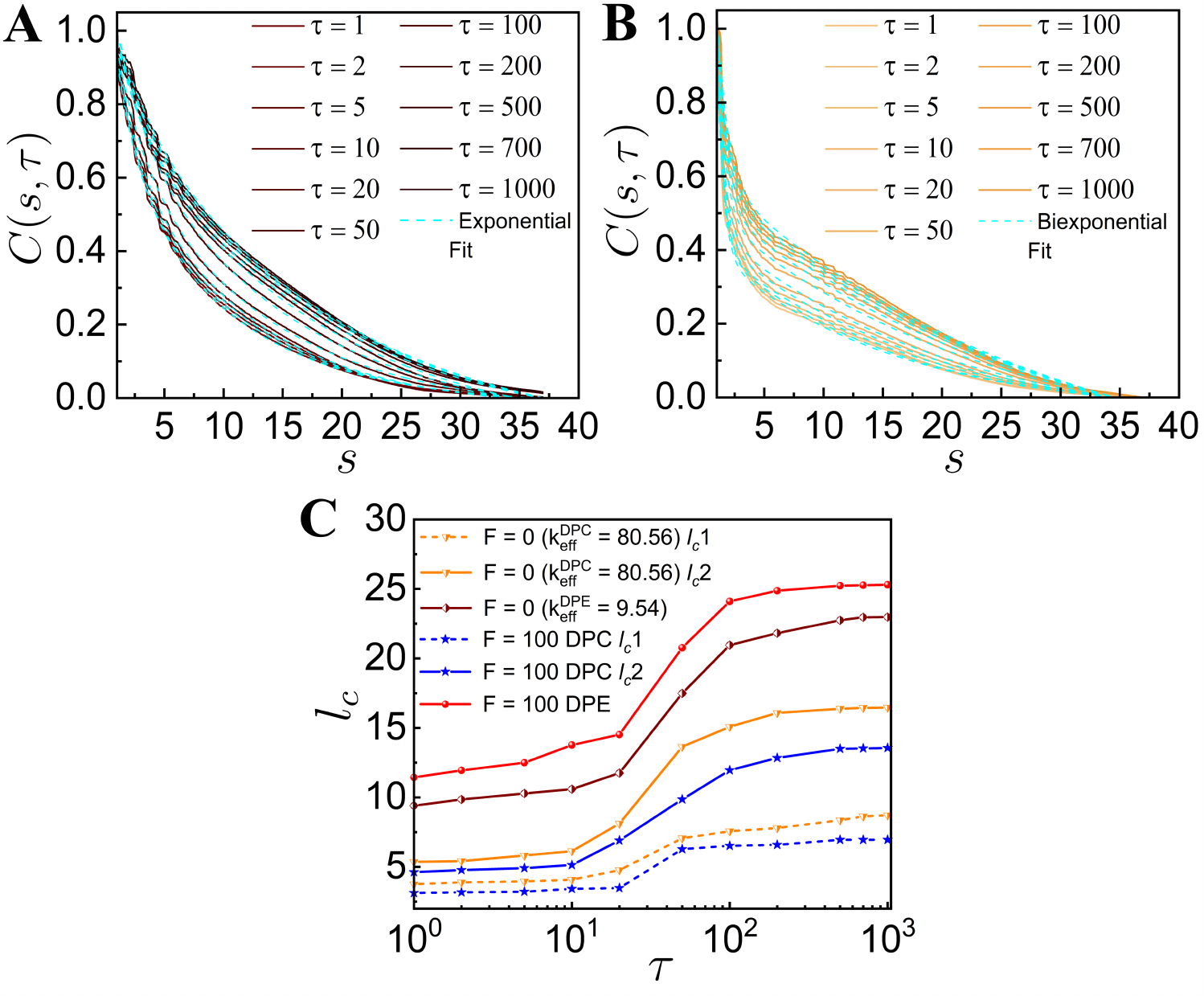} 
\caption{Spatial correlation functions $C(s,\tau)$ at different time lags, $\tau$ for the thermal chain with renormalized spring constants, $k^{\textrm{DPE}}_{\textrm{eff}}=9.54$ (A) and, $k^{\textrm{DPC}}_{\textrm{eff}}=80.56$ (B). (C) Correlation lengths ($l_c$) as a function of $\tau$ for the thermal chain with renormalized  spring constants, $k^{\textrm{DPE}}_{\textrm{eff}}$ and $k^{\textrm{DPC}}_{\textrm{eff}}$  and large DPE and DPC dipolar active forces.}\label{fig:spatial_modk}
\end{figure*}

\section{Comparison with the Effective Equilibrium Model}

\noindent In our model of dipolar activity, DPE and DPC forces are working on every bead along the unit bond vector with constant magnitude, $F$. Another option is to introduce thermal extensile  or contractile forces acting along the bond vector. The thermal extensile forces on the $n$th bead due to $(n-1)$th and $(n + 1)$th beads become $\frac{F}{\left<l_b\right>} \left(\vec{r}_n-\vec{r}_{n-1}\right)$ and $\frac{F}{\left<l_b\right>} \left(\vec{r}_n-\vec{r}_{n+1}\right)$ respectively where $\left<l_b\right>$ is the average bond length. The net force on $n$th bead, $\frac{F}{\left<l_b\right>} \left(2\vec{r}_n-\vec{r}_{n+1}-\vec{r}_{n+1}\right)$ becomes $-\frac{F}{\left<l_b\right>} \frac{\partial^2{{\vec{r}_{n}}(t)}}{\partial{n^2}}$ in  the continuum limit. Thus, the effective spring constant $k^{\textrm{DPE}}_{\textrm{eff}}$ is reduced by $\left(k-\frac{F}{\left<l_b\right>}\right)$ for the thermal extensile case. Similarly, for the thermal contractile case, $k^{\textrm{DPC}}_{\textrm{eff}}$  is enhanced by $\left(k+\frac{F}{\left<l_b\right>}\right)$. We calculate $\left<l_b\right>$ from the distribution of $P(l_b)$ (Fig. \ref{fig:MSD_bond_length_schematic}D) for both the extensile and contractile forces. Next, we simulate Eq. (\ref{eq:langevineq}) with ${\vec F}_{a,n}=0$ and modified spring constants, $k^{\textrm{DPE}}_{\textrm{eff}}=\left(k-\frac{F}{\left<l_b\right>}\right)$ and $k^{\textrm{DPC}}_{\textrm{eff}}=\left(k+\frac{F}{\left<l_b\right>}\right)$ respectively.   For $F=100$, the spring constant, $k=30$ of the thermal chain is modified to $k^{\textrm{DPE}}_{\textrm{eff}}=9.54$ for extensile chain with $\left<l_b\right>=4.88$ and to  $k^{\textrm{DPC}}_{\textrm{eff}}=80.56$ for contractile chain with $\left<l_b\right>=1.97$. We compare the results, $\left<\overline{\Delta r_n^2(\tau)}\right>$ (Fig. \ref{fig:msd_bond_modk}A), $P(l_b)$ (Fig. \ref{fig:msd_bond_modk}B) and $C(s,\tau)$ (Fig. \ref{fig:spatial_modk}) of our model of dipolar activity to thermal chain with these modified spring constants $k^{\textrm{DPE}}_{\textrm{eff}}$ and $k^{\textrm{DPC}}_{\textrm{eff}}$. We find that the thermal chain with $k^{\textrm{DPE}}_{\textrm{eff}}=9.54$ is less stretched than the chain with active dipolar extensile force $F=100$. Similarly, the thermal chain with $k^{\textrm{DPC}}_{\textrm{eff}}=80.56$ is less compressed than the chain with active dipolar contractile force $F=100$. In a semiflexible polymer, stretching and bending coefficients are not independent of each other \cite{ghosh2014dynamics}. Hence, if the stretching coefficient is modified, it also modifies the bending coefficient as evident in Fig. \ref{fig:bending_keff}B. Large DPE (or DPC) activity strongly increases (or decreases) the probability of long straightened segments as compared to thermal chain with renormalized spring constants, $k^{\textrm{DPE}}_{\textrm{eff}}$ and $k^{\textrm{DPC}}_{\textrm{eff}}$. The behavior of   steady state $\left<v^2\right>$ and PE  (see supplementary Fig. S5), $P(R_g)$  (see supplementary Fig. S6) and, $P(N)$  (see supplementary Fig. S7) for both the  dipolar activities and the thermal chain with $k^{\textrm{DPE}}_{\textrm{eff}}$ and $k^{\textrm{DPC}}_{\textrm{eff}}$ are given in the Supporting Material and we observe similar trends, DPE (or DPC) active forces promote larger (or smaller) displacements of the system than the thermal one with $k^{\textrm{DPE}}_{\textrm{eff}}$ (or $k^{\textrm{DPC}}_{\textrm{eff}}$). The stretching (or compression) of the thermal chain with lower (or higher) spring  constants as compared to the chain with large dipolar extensile (or contractile) activities arise because the magnitude of the dipolar active force is constant whereas the magnitude of the spring force is fluctuating. This leads us to propose that the extensile  (or contractile) dipolar active system can not be mapped to a thermal system with larger (or lower) bending modulus and longer (or shorter) bond lengths. \\

\noindent To test the robustness of our model of dipolar activity, we study the dynamics of a tagged monomer and the spatial correlation length of the polymer chain under spherical confinement of radius $35\sigma$. The trends of $\left<\overline{\Delta r_n^2(\tau)}\right>$ and $l_c (\tau)$ with the spherical confinement (see supplementary Fig. S9 and S10) are qualitatively similar with or without the spherical confinement.  This suggests the spatial coherence in polymer primarily arises due to the correlated motion of the beads. However, the dynamics of the tagged monomer (see supplementary Fig. S9) and the COM (supplementary Sec. II and Fig. S2). are suppressed due to the confinement. Accordingly the spatial correlation length also decreases in the presence of confinement (see supplementary Fig. S10). \\

\section{Conclusion}

\noindent Active forces inside a nucleus play an important role in both genome organization and dynamics. Experiments suggest that interphase chromatin is subjected to ATP-powered enzymes such as RNA polymerase \cite{zidovska2013micron}. Motivated by that, we implement the simplest two types of dipolar active forces, namely extensile and contractile on semiflexible, self-avoiding polymer chain. The main findings from our model are the following,
\\
\\
(1)  The MSD of a tagged monomer $\left<\overline{\Delta r_n^2(\tau)}\right>$ for DPE system shows enhanced subdiffusion than the DPC one which is related to the shifting of peak values of distribution $P(l_b)$ at large values of bond length $l_b$ in the DPE case. In our model, spatial coherence emerges with or without dipolar activity. However, the correlation length increases or decreases for the DPE or DPC system as compared  to the case with no dipolar forces. 
\\
\\
(2) Although we do not consider the hydrodynamic interactions, we show that how the local motion (related to tagged monomer’s dynamics) is associated with the large scale spatially coherent motion. Hence, the coupling between activity and elasticity (related to bending and stretching in our single polymer chain) generates the large scale spatially coherent motion. In other words, hydrodynamics is not essential to show large scale spatially coherent motion of chromatin. A simpler model like ours can capture coherent motion.
\\
\\
(3) In our model, we observe that the center of mass motion is not enhanced in the presence of dipolar active forces. This suggests that in this context, the effective temperature is not a good descriptor for dipolar activity. In that way, our model differs from typical active systems where one observes superdiffusion \cite{osmanovic2017dynamics}.
\\
\\
(4) Our work demonstrates that spherical confinement does not affect the qualitative trends of the spatial coherence. The dynamics of the tagged monomer is suppressed due to the confinement. Accordingly the spatial correlation length also decreases in the presence of confinement. The spatial coherence emerges from the correlated motion of the monomers.
\\
\\
\\
(5) More importantly, we explicitly show that the extensile or contractile dipolar active system can not be mapped to a thermal system with renormalized elasticity.  Thus, chromatin models  based on purely equilibrium forces arising from polymer elasticity and interactions \cite{di2018anomalous} can not account for the active motorized systems \cite{goychuk2014molecular}. However, our model of renormalized elasticity should be verified with other values of $k_{\textrm{eff}}$.
\\
\\
\noindent A recent Optical experiment showed that chromatin movement is coherent over a length scale of  $4-5 $ $\mu m$ and persists for $5-10$ seconds \cite{zidovska2013micron}. In order to make a comparison with experiment, we consider each monomer represents $50000$ base pairs. Therefore, we assume the diameter of each monomer is approximately $150$ nm \cite{liu2018chain,di2018anomalous}. Considering the nuclear viscosity $\eta = 7$ cP \cite{hajjoul2013high} and $T=298K$, the Brownian time scale of single particle is coming out to be $\tau_0=\frac{3 \pi \eta \sigma^3}{k_B T} \approx 54$ ms. From Fig. \ref{fig:spatial} F, one can see that for the extensile force of magnitude $F = \frac{100 k_B T}{\sigma} \approx 2.74$ pN, the correlation length saturates at $l_c \approx 24 \sigma$ $\approx 3.6$ $\mu m$ and persists for $100 - 1000 \tau_0$ $\approx 5.4-54$ seconds. \\

\noindent The physics underlying the phenomena we report here relies on the motion of a long semiflexible polymer driven by the  dipolar active forces. We do not take into account the full complexity of the chromatin packing \cite{langowski2007computational, fritsch2011chromosome} or any salt-induced long-ranged electrostatic interactions \cite{cherstvy2006layering}. From our work, it is however clear that the coupling between activity and elasticity (related to bending and stretching of our single polymer chain) is sufficient to qualitatively explain the experimentally observed large scale spatially coherent motion of  chromatin.  We hope that the simplicity of our model helps us understand the structural and dynamical behavior of the energy consuming processes in chromatin and allows one to add more complexity to the problem, such as  nuclear lamina. \\

\section*{Acknowledgments}
 \noindent We thank Nir Gov and Guang Shi for many stimulating discussions. S.C. thanks DST Inspire for a fellowship. L.T. thanks UGC for a fellowship. R.C. acknowledges SERB for funding (Project No. MTR/2020/000230 under MATRICS scheme). We acknowledge the SpaceTime-2 supercomputing facility at IIT Bombay for the computing time. \\

\section*{Supplemental Material}

\renewcommand{\thesubsection}{\Roman{subsection}}
\setcounter{subsection}{0}
\section{Mean Square Displacement of the center of mass}

\noindent The most straightforward tool for understanding the properties of motion from the trajectories is the mean square displacement (MSD). The time-averaged MSD of the center of mass (COM) is defined as $\overline{\Delta r_c^{2}(\tau)} = \frac{1}{T-\tau} \int_{0}^{T-\tau} {\left[ \vec{r}_c(t_0+\tau) - \vec{r}_c(t_0)\right]}^2  dt_0$ from the time series of the position of the COM  $\vec{r}_c(t)$ where $T$ is the total run time and $\tau$ is the lag time. The time-and-ensemble-averaged MSD of the COM is obtained as:  $\left\langle{\overline{\Delta r_c^{2}(\tau)}}\right\rangle  =  \frac{1}{N^\prime} \sum_{i=1}^{N^\prime}{\overline{\Delta r_c^{2}(\tau)}}$  where $N^\prime$ is the number of independent trajectories. The MSD $\left(\left<\overline{\Delta r_c^2(\tau)}\right>\right)$ of the center of mass (COM) diffuses freely at all times in the presence or absence of extensile and contractile activity (Fig. \ref{fig:msd_com}). In addition, all the active and passive $\left<\overline{\Delta r_c^2(\tau)}\right>$ are merging with each other (Fig. \ref{fig:msd_com}). However, in the long time limit, the dynamics of the center of mass of the polymer chain will be obstructed by the spherical confinement (Fig. \ref{fig:msd_confinement}).

\section{Distribution functions of the radius of gyration}

\noindent To further quantify the effects of the contractile and extensile active forces on the loci mobilities, we analyze the probability distribution, $(P(R_g))$ of the radius of gyration $R_g$ of the chain where $R_{g} = \sqrt{ \frac{1}{\text{N}}  \sum_{i=1}^{N}  \left (r_i - r_{\text{c}} \right)^2}$. For smaller DPC and DPE activities, $P(R_g)$s are Gaussian and appear to be qualitatively  similar to $P(R_g)$ of passive system. But, $P(R_g)$s for high DPE and DPC forces show well-separated peaks (Fig.  \ref{fig:histogram_gyration}). For large DPC force, $P(R_g)$ is Gaussian and sharply peaked at lower values of $R_g$. The broader distribution of $P(R_g)$ with peak at higher value of $R_g$ deviates from the Gaussian  for large DPE force which accounts for the enhancement of stretching of the polymer chain.  For the high DPC force, the deviation of the peak value of $P(R_g)$ from the passive case  is small as compared to the deviation for the DPE force. This small deviation for the DPC force is most probably due to the self-avoidance of the beads. To test the effect of self-avoidance as an energetic barrier, we compute the potential energy and the mean square fluctuation of the velocity $(v)$, $\left<v^2\right>$ per particle of the polymer chain at each instant of time in the steady state. The velocity of a monomer over the integration time step $\delta t$ is defined as $v=\frac{\vec{r}_{n}(t+\delta t)-\vec{r}_{n} (t)}{\delta t}$. It is a quantity that can be measured experimentally by tracking the position of a chromosomal locus as a function of time \cite{lampo2016physical} and has been extensively used to  determine the nature of the stochastic process underlying the observed motion \cite{put2019active,di2018anomalous}. PE per particle increases more for the DPC forces than for the DPE forces (Fig. \ref{fig:energy}A) as the effect of self-avoidance translates to an increasing potential energy barrier for the DPC forces. On the other hand, $\left<v^2\right>$ per particle increases more for the DPE forces than for the contractile forces (Fig. \ref{fig:energy}B) as the DPE forces show larger displacements than the DPC forces.

\section{Number fluctuations}

\noindent To gain a deeper understanding of the underlying complex movement of the monomers, we investigate the number fluctuations. To compute the number fluctuations, we randomly select a number of small spherical probes of radius $4\sigma$ at different locations within the system.  We then construct a histogram over different spherical probes and independent trajectories from which probabilities $(P(N))$ are computed. In equilibrium, the statistics of fluctuations in $N$ inside a small spherical probe volume follows Gaussian distribution. For small values of DPE and DPC forces, we find that the the probability distributions $P(N)$ are indeed Gaussian (Fig. \ref{fig:density}). For large DPC force, $P(N)$ also exhibits the Gaussian distribution. For large DPC force, the peak of $P(N)$ shifts towards a higher value of $N$ and the width of $P(N)$ becomes narrower than the passive one (Fig. \ref{fig:density}). Hence, large DPC force generates high density regions, leading to decrease in mobility. On contrary, for large DPE force, $P(N)$ decays exponentially with $N,$ indicating that  DPE forces promote larger displacements of the system than the passive one. \\

\renewcommand{\thetable}{S\arabic{table}}
\setcounter{table}{0}
\begin{table}[h]
\centering
\begin{tabular}{ | m{8em} | m{1.5cm}| m{1.5cm} | m{1.5cm} |m{1.5cm} |m{1.5cm} |} 
\hline
$F$ & $\alpha$ & $\beta$ \\ 
\hline
$0$ & $0.88$ & $0.77$ \\ 
\hline
$4$ (DPE) & $0.88$ & $0.65$ \\ 
\hline
$4$ (DPC)  & $0.89$ & $0.63$ \\
\hline
$100$ (DPE)  & $0.93$ & $0.68$ \\
\hline
$100$ (DPC)  & $0.74$ & $0.67$ \\
\hline
\end{tabular}
\caption{The exponents, $\alpha$ and $\beta$ of $\left<\overline{\Delta r_n^2(\tau)}\right>$ at short and long times for the passive and different dipolar activities.}
\label{tab:exponent}
\end{table}

%\clearpage
\renewcommand{\thefigure}{S\arabic{figure}}
\setcounter{figure}{0}
\begin{figure}
        \centering
        \includegraphics[width=0.95\linewidth]{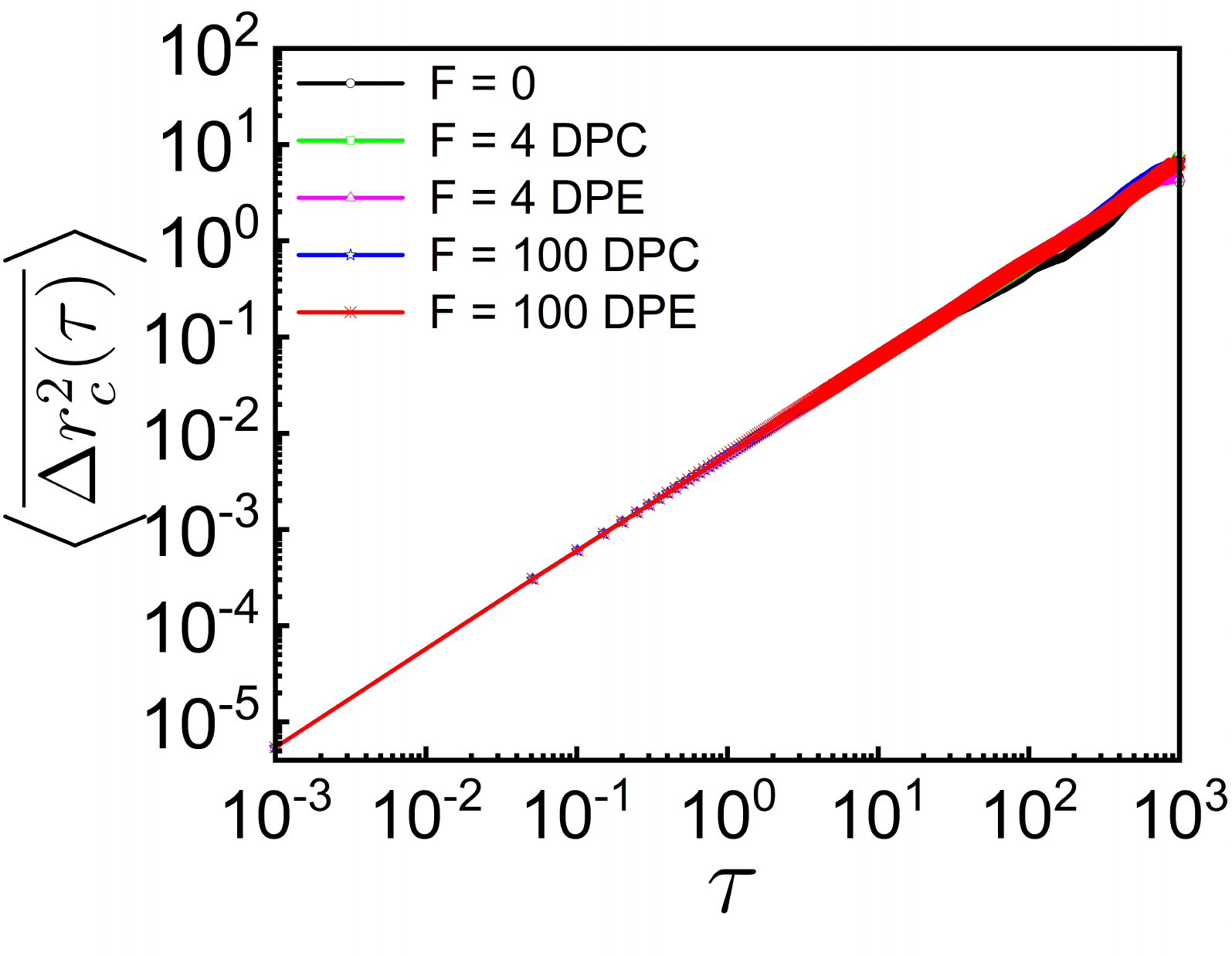}
        \caption{Log-Log plot of MSDs $\left(\left<\overline{\Delta r_c^2(\tau)}\right>\right)$ as a function of lag time $\tau$ at different DPE and DPC active forces for the COM of polymer chain.}\label{fig:msd_com}
\end{figure}

\begin{figure}
        \centering
        \includegraphics[width=0.95\linewidth]{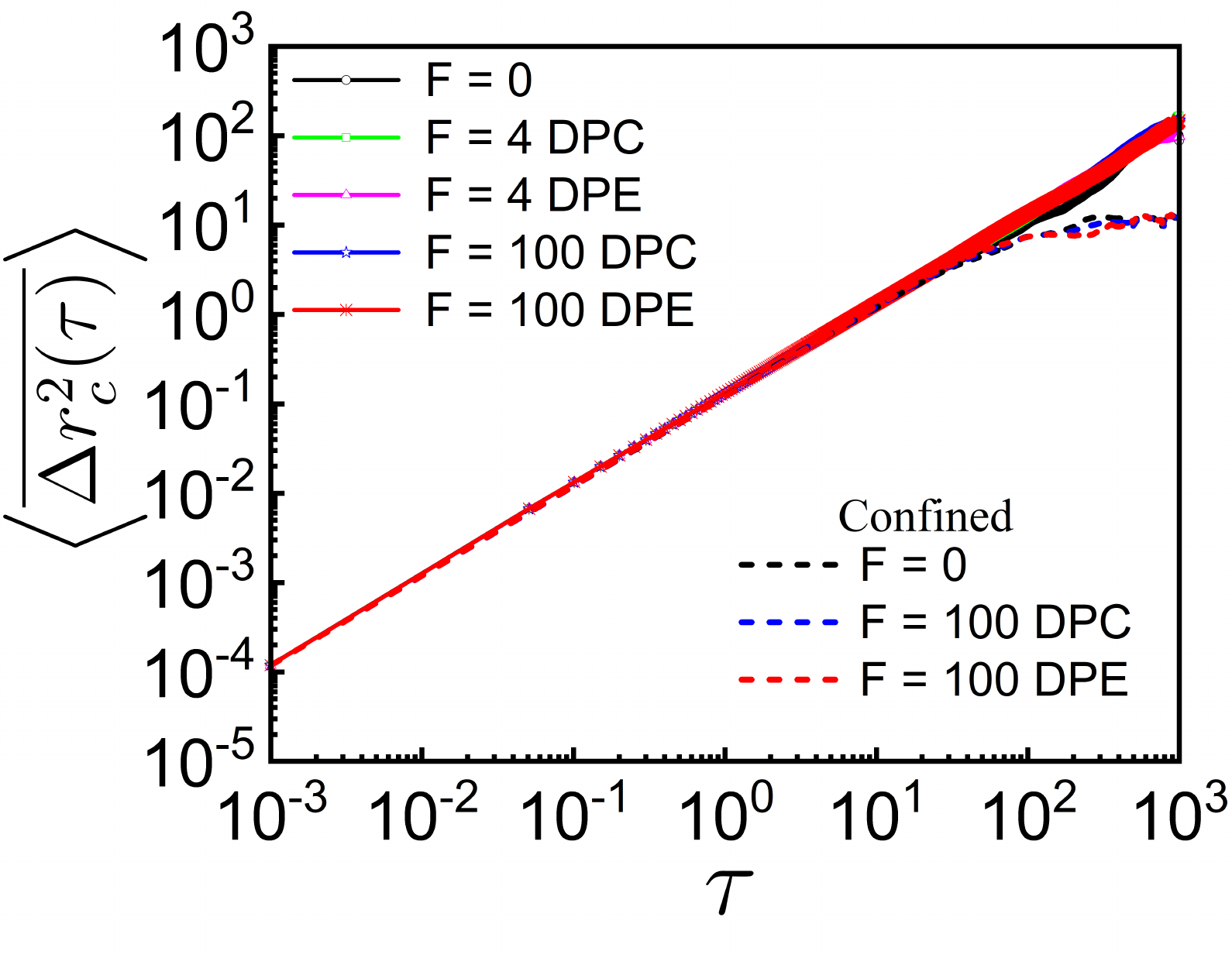}
        \caption{Log-Log plot of MSDs $\left(\left<\overline{\Delta r_c^2(\tau)}\right>\right)$ as a function of lag time $\tau$ at different DPE and DPC active forces for the COM of polymer chain under spherical confinement.}\label{fig:msd_confinement}
\end{figure}

\begin{figure*}
	\centering
	\begin{tabular}{cc}
	        \includegraphics[width=0.45\linewidth]{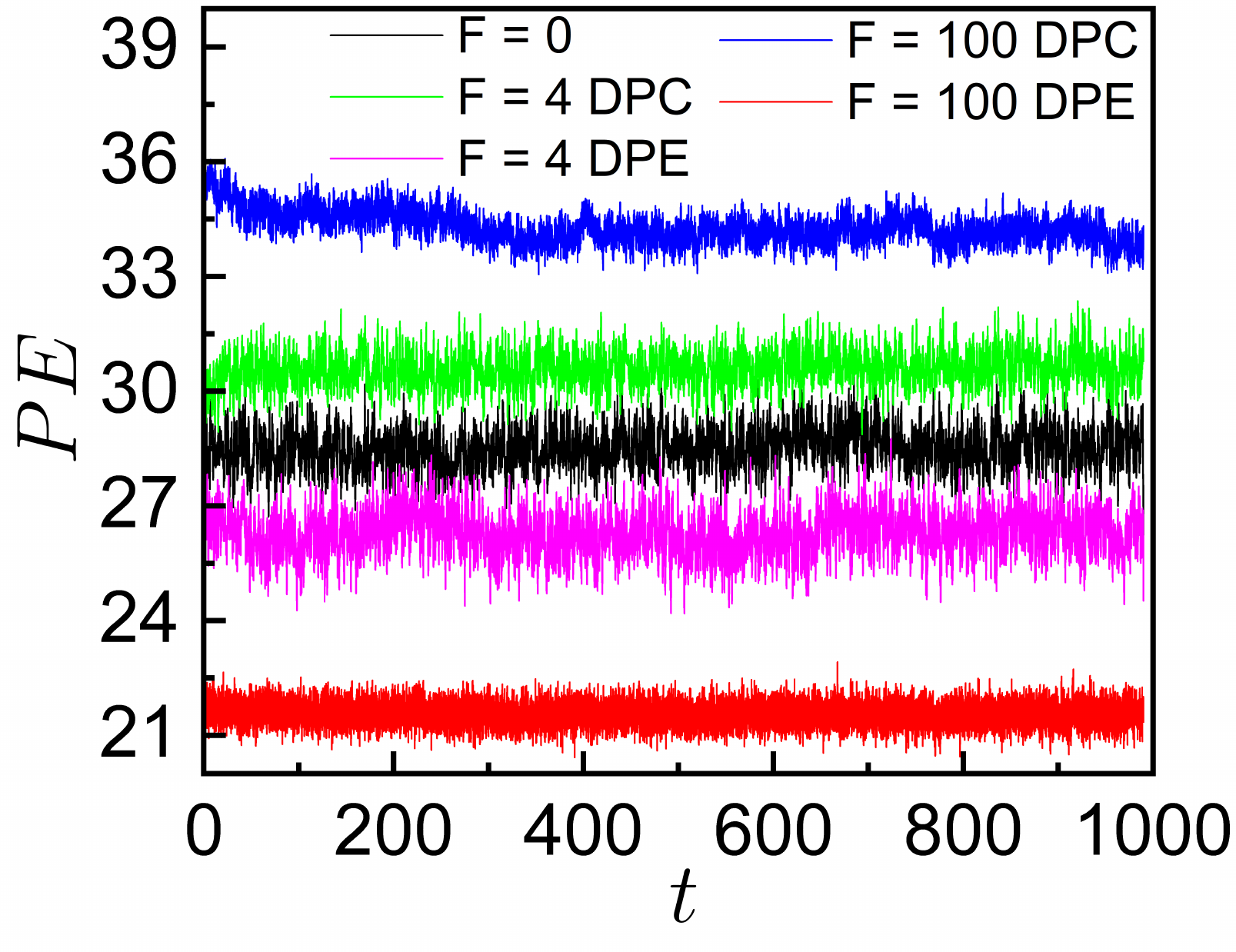} &
		\includegraphics[width=0.45\linewidth]{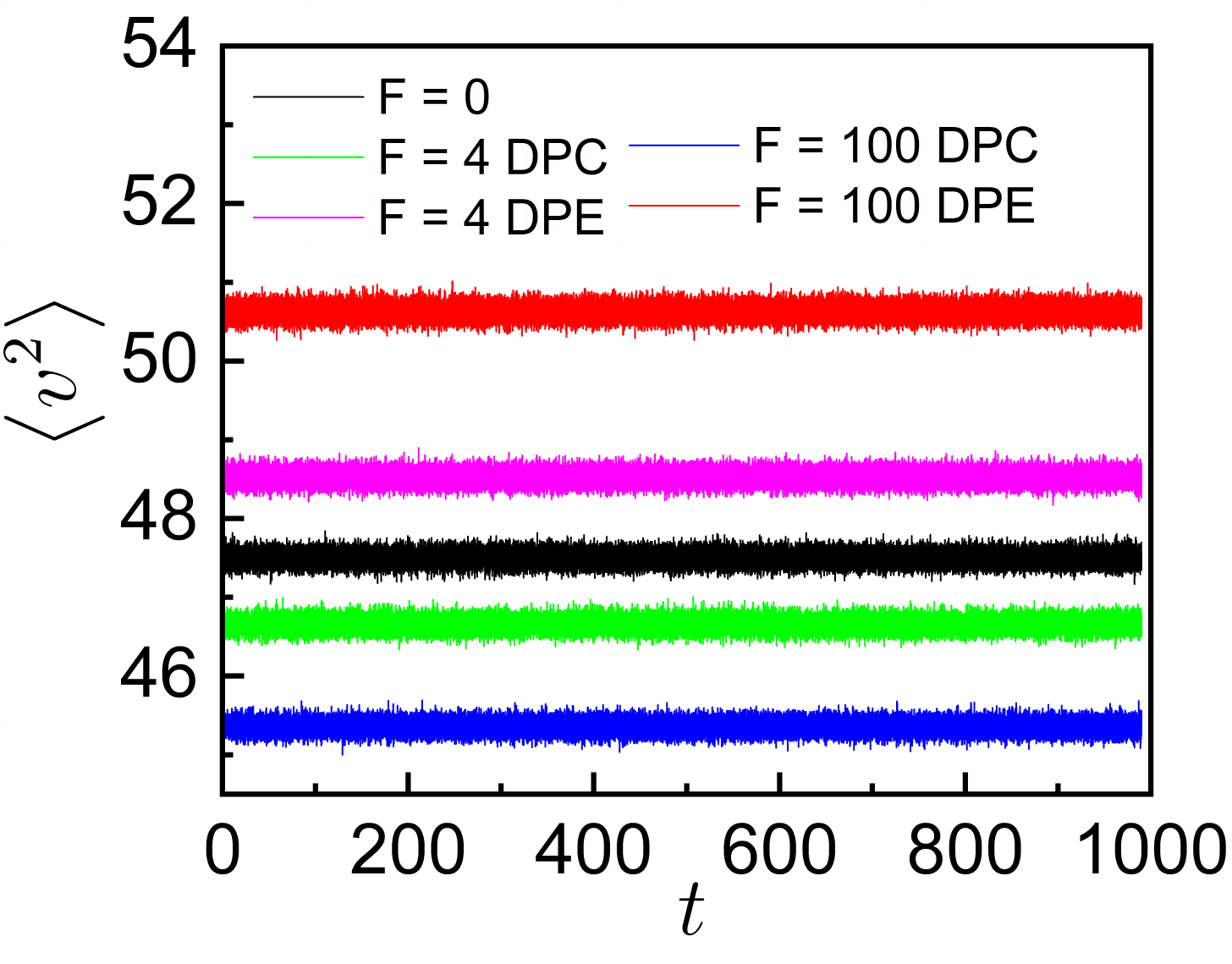}\\
		(A) & (B) \\
	\end{tabular}
\caption{Steady state PE per particle (A) and $\left<v^2\right>$ per particle (B) as a function of time $t$ at different DPE and DPC active forces for polymer chain.}\label{fig:energy}
\end{figure*}

\begin{figure*}
        \centering
        \includegraphics[width=0.5\linewidth]{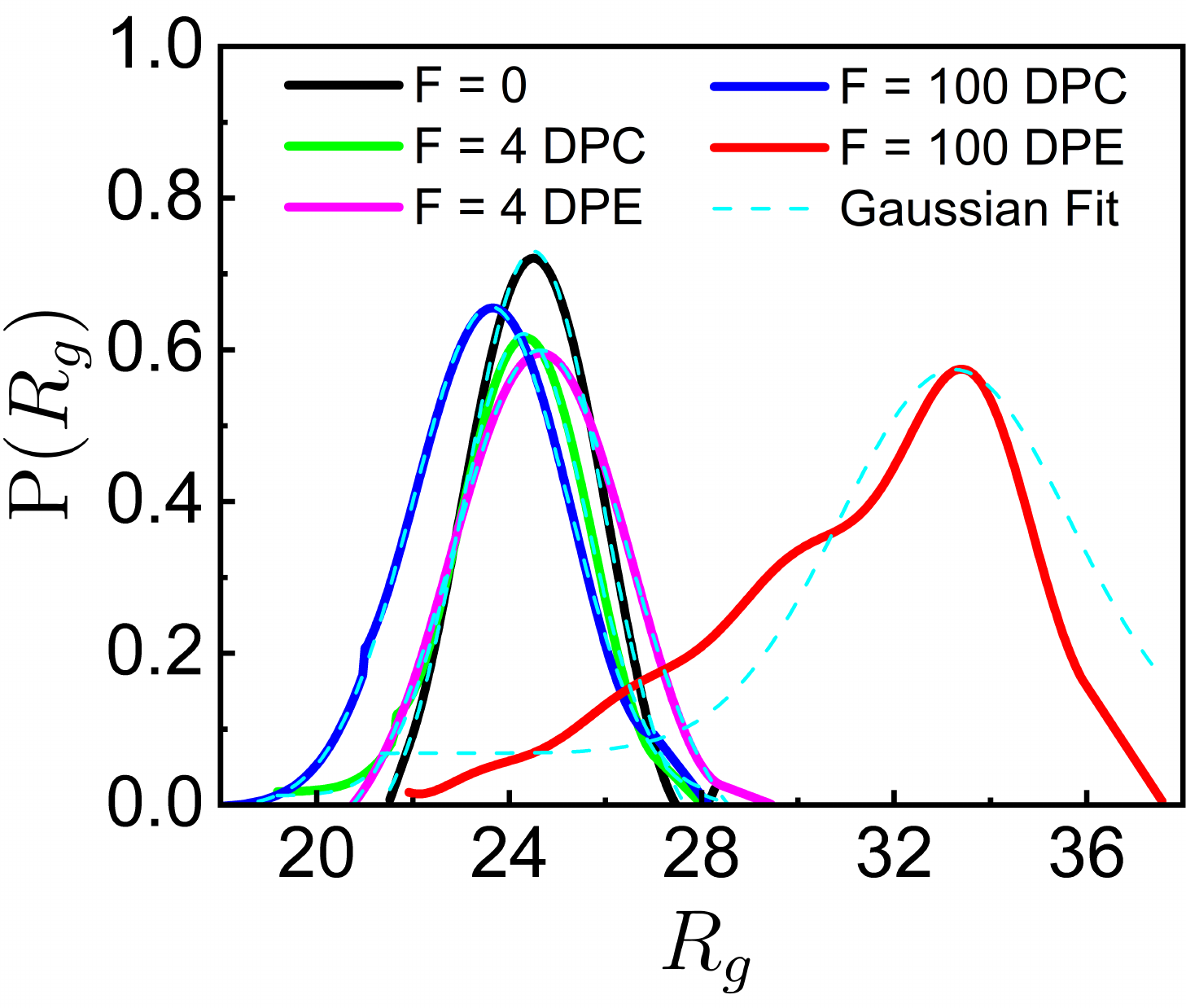}
        \caption{The probability distribution function $(P(R_g))$ of the radius of gyration $(R_g)$ in the presence of  at different DPE and DPC active forces.}\label{fig:histogram_gyration}
\end{figure*}

\begin{figure*}
        \centering
        \includegraphics[width=0.5\linewidth]{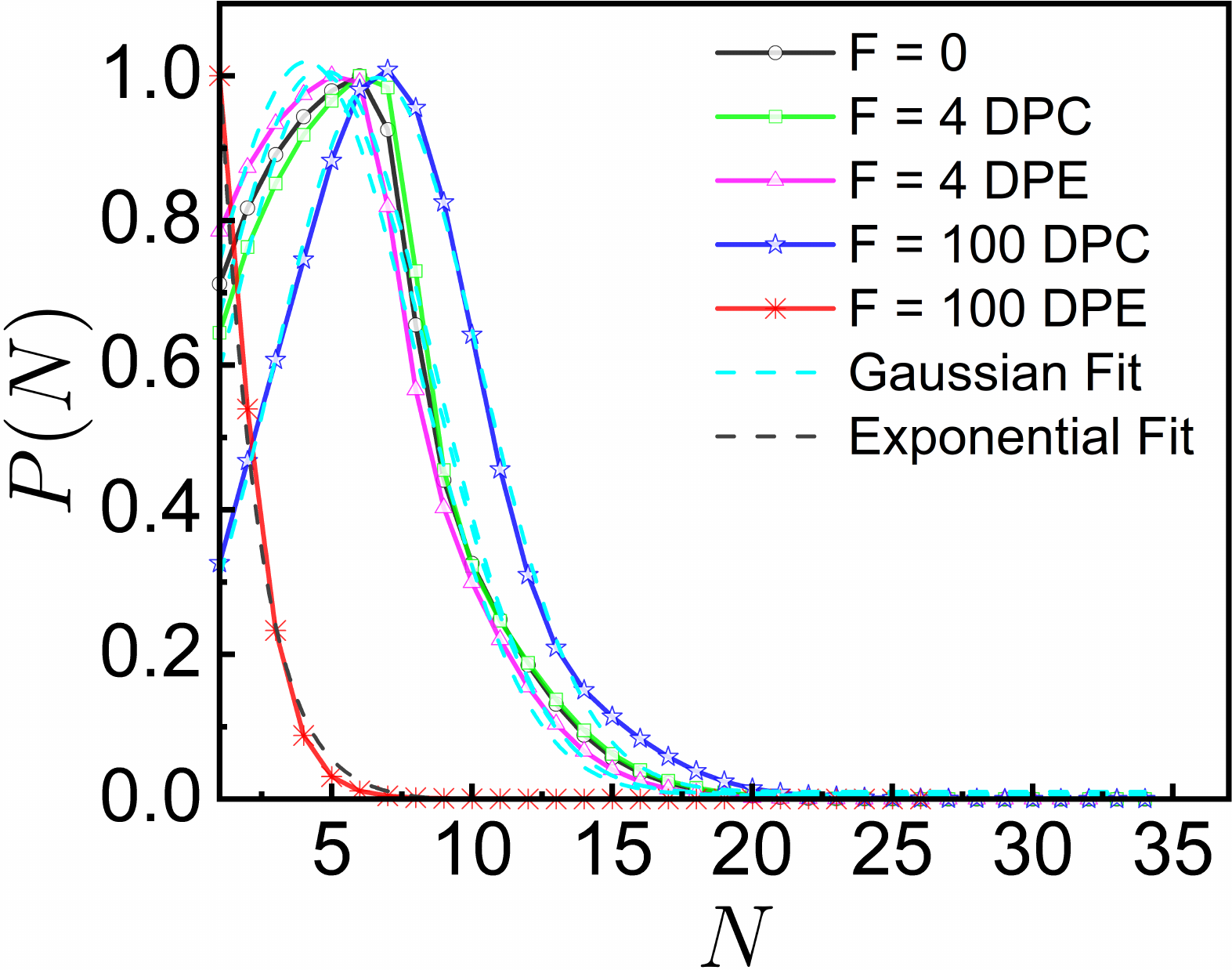}
        \caption{Number density fluctuations $(P(N))$ at different DPE and DPC activities..}\label{fig:density}
\end{figure*}

\begin{figure*}
	\centering
	\begin{tabular}{cc}
	        \includegraphics[width=0.45\linewidth]{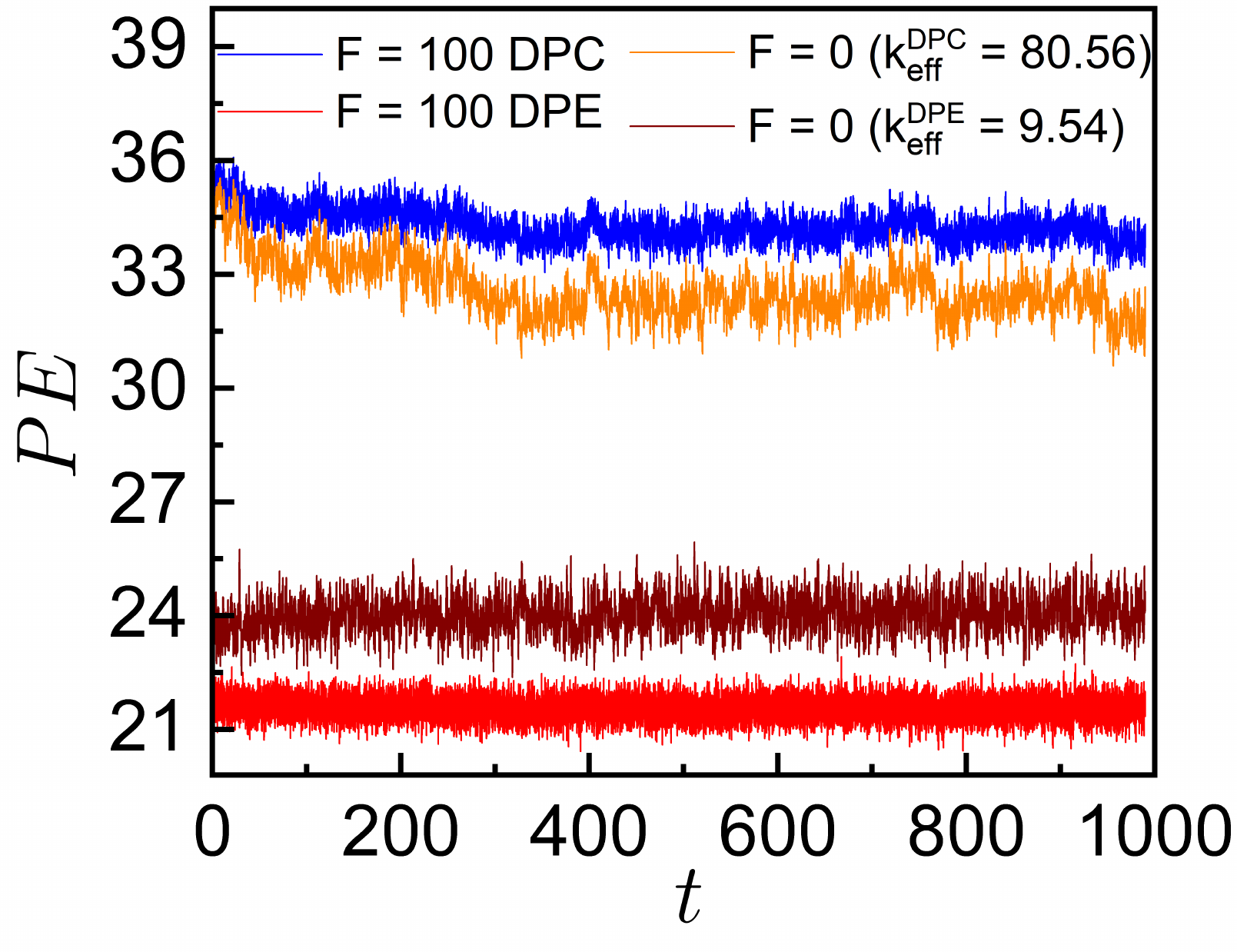} &
		\includegraphics[width=0.45\linewidth]{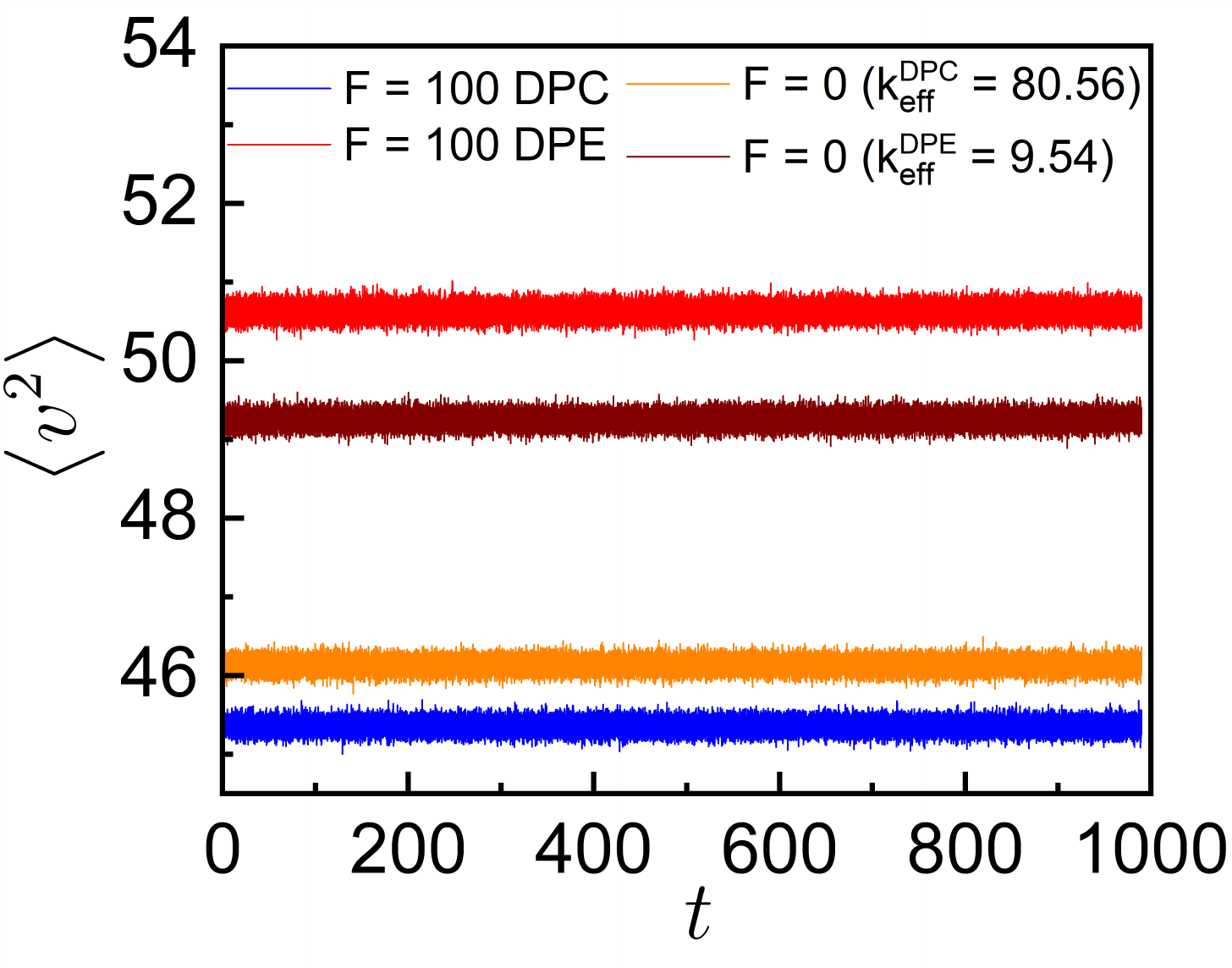}\\
		(A) & (B) \\
	\end{tabular}
\caption{Steady state PE per particle (A) and $\left<v^2\right>$ per particle (B) as a function of time $t$ for the thermal chain with effective spring constants $k^{\textrm{DPE}}_{\textrm{eff}}$ and $k^{\textrm{DPC}}_{\textrm{eff}}$  and large DPE and DPC dipolar active forces.}\label{fig:energy_modk}
\end{figure*}

\begin{figure*}
        \centering
        \includegraphics[width=0.5\linewidth]{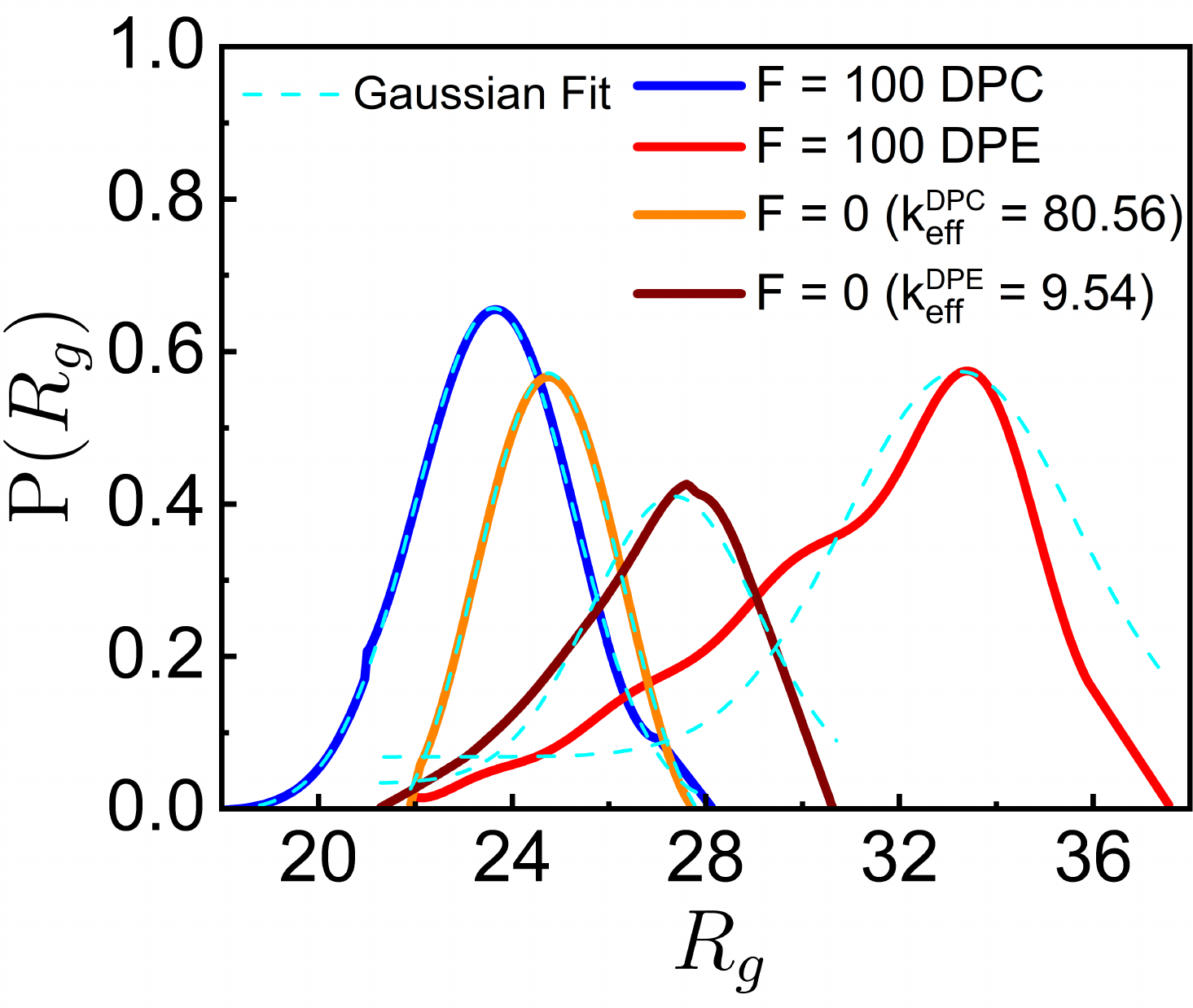}
        \caption{The probability distribution function $(P(R_g))$ of the radius of gyration $(R_g)$ for the thermal chain with effective spring constants $k^{\textrm{DPE}}_{\textrm{eff}}$ and $k^{\textrm{DPC}}_{\textrm{eff}}$  and large DPE and DPC dipolar active forces.}\label{fig:histogram_gyration_modk}
\end{figure*}

\begin{figure*}
        \centering
        \includegraphics[width=0.5\linewidth]{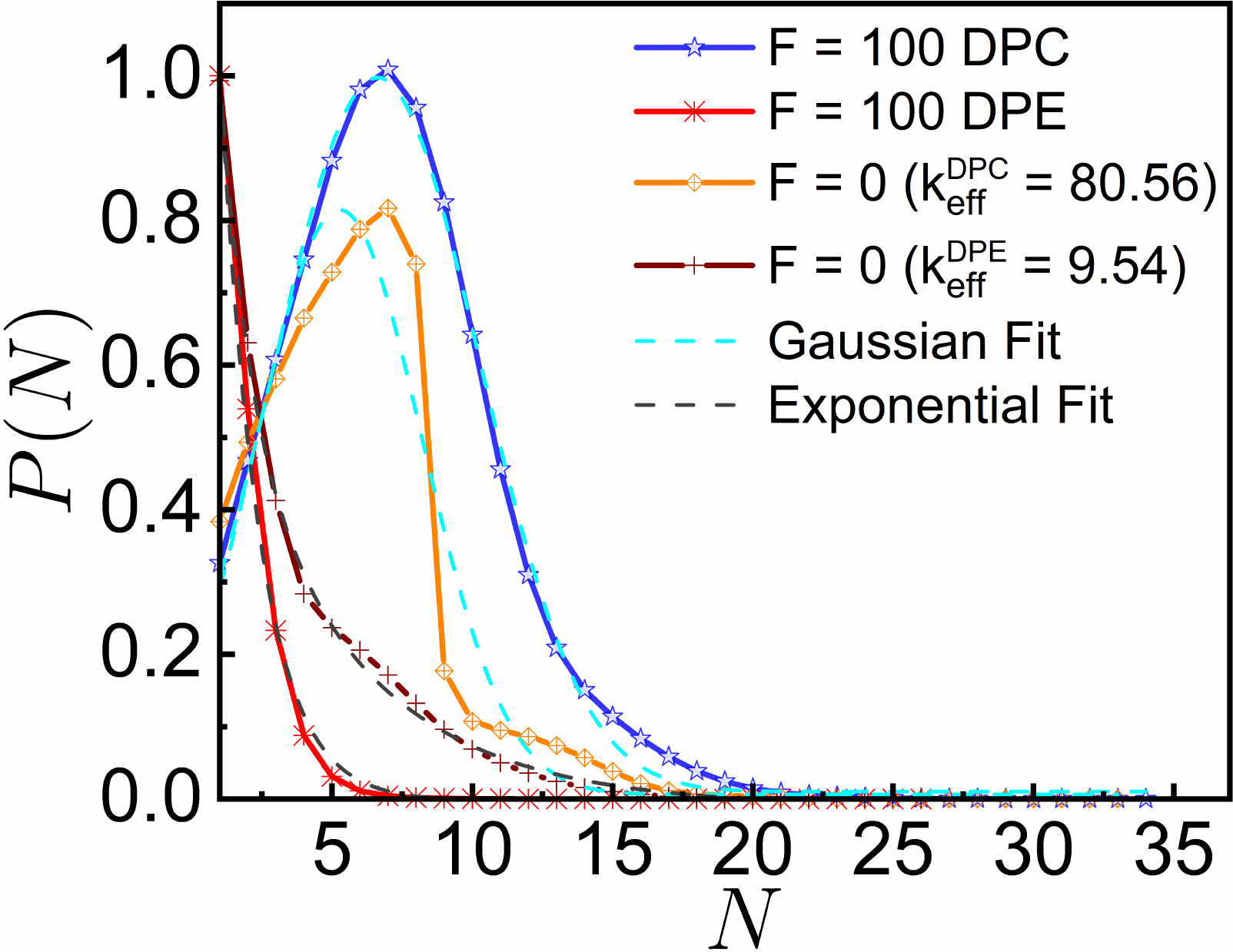}
        \caption{Number density fluctuations $(P(N))$ for the thermal chain with effective spring constants $k^{\textrm{DPE}}_{\textrm{eff}}$ and $k^{\textrm{DPC}}_{\textrm{eff}}$  and large DPE and DPC dipolar active forces.}\label{fig:density_modk}
\end{figure*}

\begin{figure*}
\centering
	\includegraphics[width=0.5\linewidth]{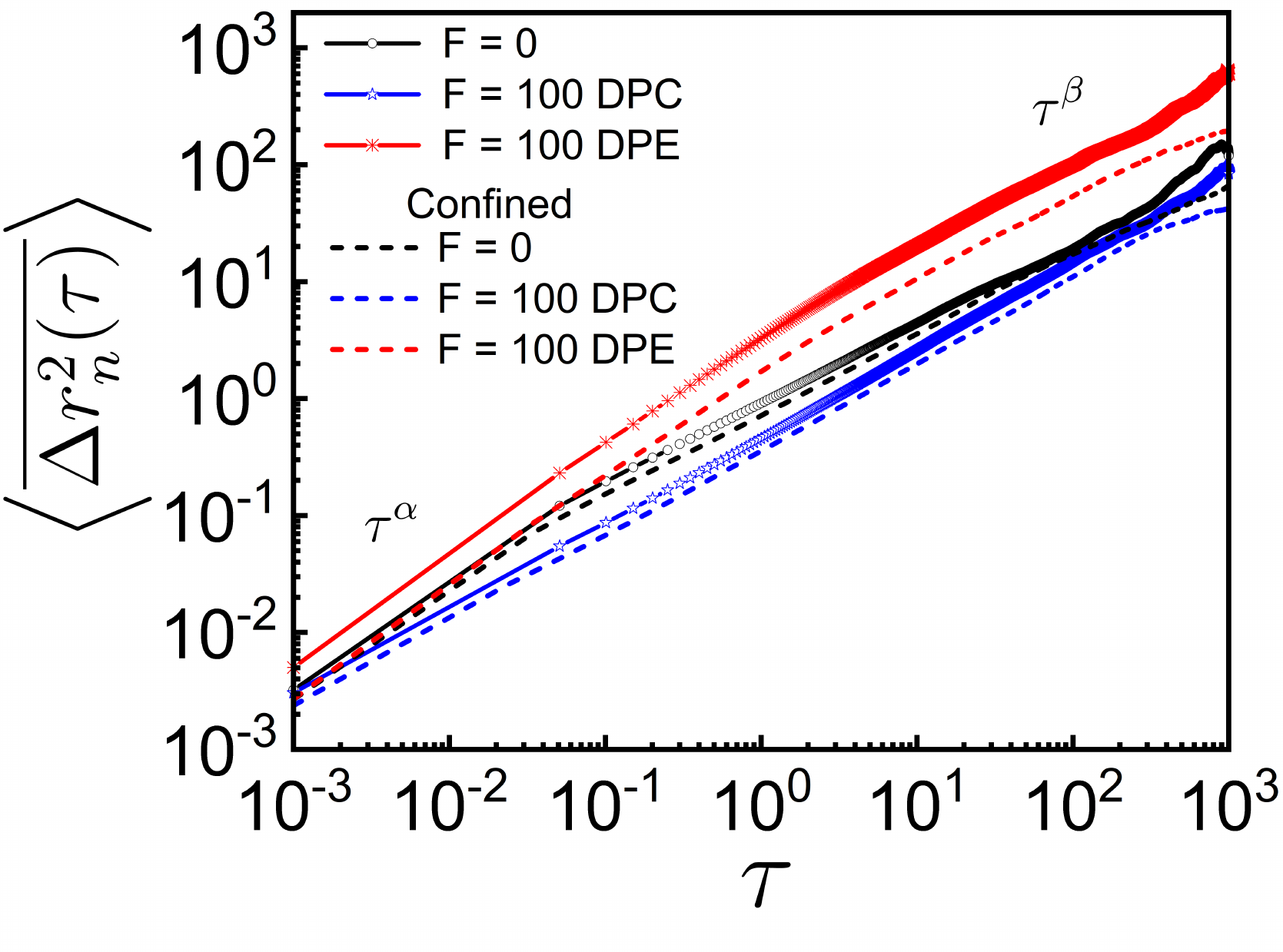} 
\caption{ Log-Log plot of tagged monomer MSDs $\left(\left<\overline{\Delta r_n^2(\tau)}\right>\right)$ as a function of lag time $\tau$ for the passive, DPE and DPC activities with (dotted lines) and without (solid lines) spherical confinement. }\label{fig:tagged_confinement}
\end{figure*}

\begin{figure*}
\centering
	\includegraphics[width=0.9\linewidth]{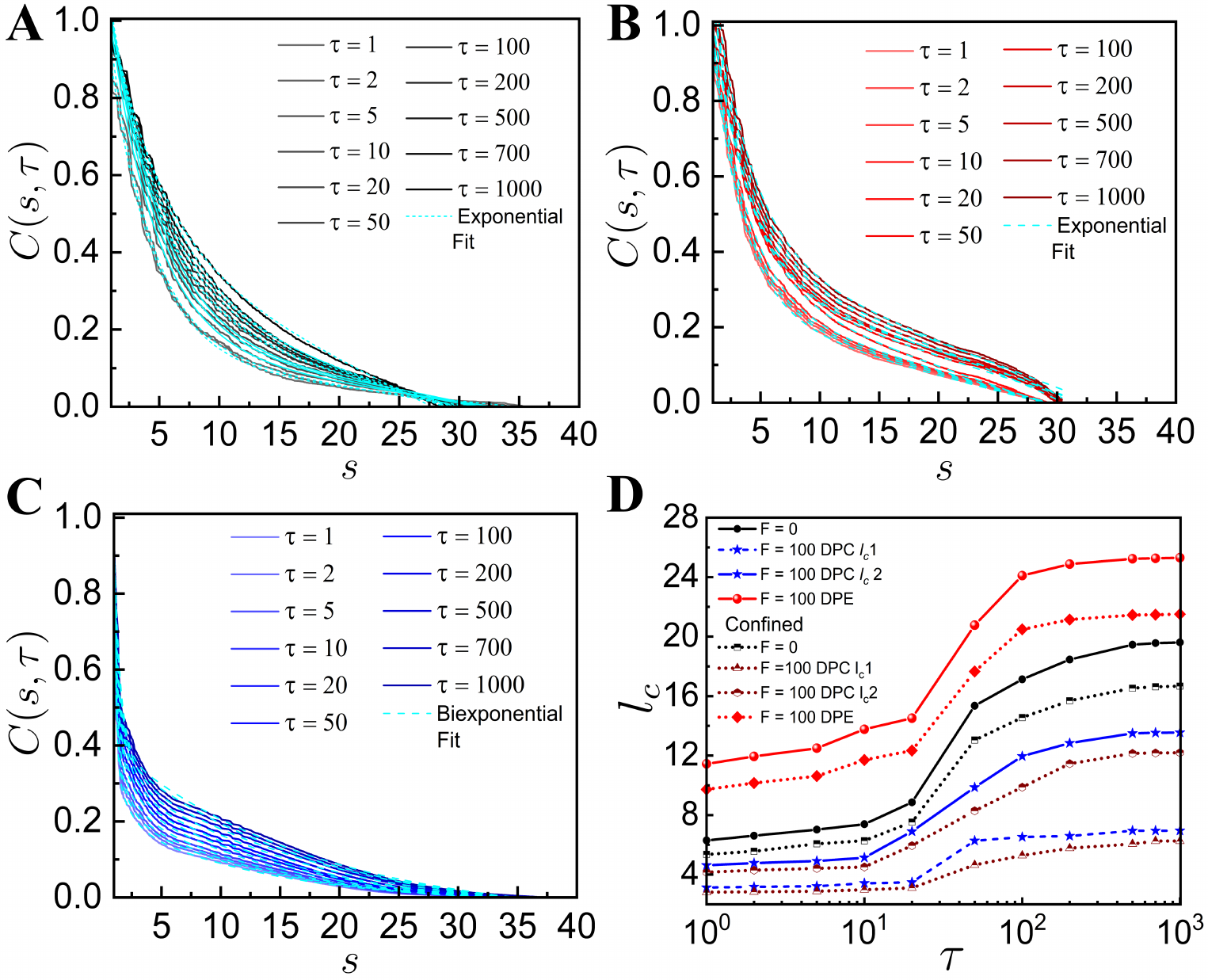} 
\caption{Spatial correlation functions $C(s,\tau)$ for the passive (A), DPE and DPC cases  for F = 100 (B and C) respectively  at different time lags, $\tau$ under sperical confinement. (D) Correlation lengths ($l_c$) as a function of $\tau$ for the passive, DPE and DPC activities  with (dotted lines) and without (solid lines) spherical confinement. }\label{fig:spatial_confinement}
\end{figure*}
\clearpage
%\bibliography{arXiv_Version_Chromatin_March2023}
%\bibliographystyle{rsc}
\providecommand*{\mcitethebibliography}{\thebibliography}
\csname @ifundefined\endcsname{endmcitethebibliography}
{\let\endmcitethebibliography\endthebibliography}{}

\end{document}